\documentclass[11pt,epsfig]{article}
\usepackage{graphicx,bm,color}
\usepackage{amsmath}
\usepackage{amssymb}
\usepackage{amsfonts}
\usepackage{epsfig}
\title{\bf Dark matter as the Bose-Einstein condensation \\in \\loop quantum cosmology}
\author{K. Atazadeh$^{1}$  \thanks{email:
atazadeh@azaruniv.ac.ir}\,\ F. Darabi$^{1,2}$\thanks{email:
f.darabi@azaruniv.edu; Corresponding author}\, and\, M. Mousavi$^{1}$\thanks{email:
mousavi@azaruniv.ac.ir}
\\ $^1${\small Department of Physics, Azarbaijan Shahid Madani University , Tabriz, 53714-161 Iran}\\$^2${\small Research Institute for Astronomy and Astrophysics of Maragha (RIAAM),
Maragha 55134-441, Iran}}

\begin{document}

\maketitle

\begin{abstract}
 We consider the FLRW universe in a loop quantum cosmological model filled with the radiation, baryonic matter (with negligible pressure), dark energy and dark matter. The dark matter sector is supposed to be of Bose-Einstein condensate type. The Bose-Einstein condensation process in a cosmological context by supposing it as an approximate first order phase transition, has been already studied in the literature. Here, we study the evolution of the physical quantities related to the early universe description such as the energy density, temperature and scale factor of the universe, before, during and after the condensation process. We also consider in detail the evolution era of the universe in a mixed normal-condensate dark matter phase. The behavior and time evolution of the condensate dark matter fraction is also analyzed.
 \end{abstract}
\vspace{2cm}

\section{Introduction}
 When one considers a universe following the standard Einstein cosmology, that is, when the dynamics of the universe is described by the general relativity equations, one deduces that there exists a mysterious singularity at the
beginning of time, so called big bang. This singularity may be considered as a deficiency of Einstein cosmology at high energies \cite{1}. Actually, one can assert that the big bang implies the breakdown of general relativity at scales with high energies, whereas we know from the observational evidences, such as the existence of the cosmic microwave background, that the big bang model works at scales lower than Planck energy scale. At these scales, the universe was full of a hot photon-baryon combination almost described by a radiation fluid. This hot combination was cooling down as the universe was experiencing the expansion,  and due to the different scaling behaviors of pure radiation and pure matter, the energy density of non-relativistic matter started to dominate over the energy density of radiation  and led to the formation of structures.

Loop quantum gravity (LQG) is one of the potential candidates for the study of quantum gravity \cite{LQG}.
Homogenous and isotropic space time, reduces Loop quantum gravity (LQG) to loop quantum cosmology (LQC) \cite{LQC}. According to this theory, due to the effective quantum gravitational effects, the big bounce happens to remove the big bang singularity, after which a super-inflation phase occurs, and
then the universe enters a normal inflation regime \cite{2}, \cite{3}. The loop quantum cosmology effects  manifest themselves in a new effective form of  modified Friedmann equation. These effective equations have been derived in the literature for FLRW loop quantum cosmologies considering different matter sectors at early universe \cite{multiple}. 

An extended scenario of the matter bounce cosmology, considering a single scalar field with an approximately exponential potential, was proposed as an alternative to the slow roll inflation, in the context of (LQC), in which the universe has experienced a quasi-matter contracting phase with a variable equation of state parameter \cite{p1}. This matter bounce scenario, in the teleparallel version of (LQC), was shown to be in good agreement with the new BICEP2 data \cite{p2}. A Gauss-Bonnet extension of (LQC), by introducing holonomy corrections in modified $f(G)$ theories of gravity was also developed where the authors have provided a perturbative expansion in the critical density as well as a parameter characteristic of (LQG), and obtained leading order corrections to the classical $f(G)$ theories of gravity. They also presented a reconstruction method which makes possible to find the (LQC) corrected $f(G)$ theory capable of realizing various cosmological scenarios \cite{p3}. In another work, in order to avoid singularities of $f(R)=R+aR^2$ model, holonomy corrections to this model were introduced in Einstein frame and a detailed analytical and numerical study was performed when holonomy corrections are taken into account in both Jordan and Einstein frames. They obtained, in Jordan frame, a dynamics which is different qualitatively, from the one of the original model, at early times. According to this dynamics, the universe is not singular, neither at early times in the contracting phase, nor at bouncing to enter in the new expanding inflationary phase. This dynamics may lead to better predictions for the inflationary phase in comparison to the current observations \cite{p4}.

Motivated by the above mentioned success of (LQC) in describing the very early stage of inflationary universe, we are encouraged to study the evolution of universe after the inflationary era and investigate the possible establishment
of  a specific kind of dark matter (see below) and study its contribution  in the subsequent
evolution of the universe. Explicitly, we assume that after inflation the universe includes ordinary dark matter (bosonic particles), dark energy, radiation and baryonic matter, and that the ordinary dark matter component in the sufficiently cooled universe has gone through a phase transition  provided  by the Bose-Einstein condensation, which might have happened during the early stages of cosmological evolution of the universe with a low temperature comparable to the critical temperature for Bose-Einstein condensation $T_{{\rm cr}} \sim 2\pi \hbar^{2}n^{2/3}/mk_{{\rm B}}$ where $m$ is the particle mass, $n$ is the particle density and $k_{{\rm B}}$ is Boltzmann's constant \cite{6}-\cite{10}.

Particularly, in the Bose-Einstein Condensation  (BEC) model, the dark matter can be described as a non-relativistic Newtonian gravitational condensate, whose pressure and energy density are related by a barotropic equation of state. Based on the above arguments, we assume the possibility that the Bose-Einstein condensation might have happened during the early stages of cosmological evolution of the universe with a temperature comparable to the critical temperature for Bose-Einstein condensation $T_{{\rm cr}} \sim 2\pi \hbar^{2}n^{2/3}/mk_{{\rm B}}$ where $m$ is the particle mass, $n$ is the particle density and $k_{{\rm B}}$ is Boltzmann's constant \cite{6}-\cite{10}.

 The modification of cosmological evolution of the early universe, in the presence of condensate dark matter in standard cosmology, has already been studied in \cite{16}. In this paper, we generalize this model from standard cosmology to LQC. The organization of the present paper is as follows. In section $2,$ we outline the modification  of Friedmann equations in LQC model and the basic properties of the normal and BEC dark matter in addition to the presentation of the relevant physical quantities and the equation of state. In section $3,$ we give a brief explanation about the cosmological dynamics of the Bose-Einstein condensate. Next, we introduce the numerical values of the cosmological parameters at the condensation point, and then obtain the important $h(t)$ parameter as a volume fraction of matter in the Bose-Einstein condensed phase which explains the evolution of the dark matter energy density during the transition process, and extract the order of BEC time interval in LQC model. In the following, we study the post condensation phase in LQC model describing the evolution of the scale factor parameter under the effect of the existence of the BEC dark matter in LQC. The paper ends with a conclusion.

{\section{ Bose-Einstein condensate dark matter in the    LQC universe}}

In this section, we intend to review the Bose-Einstein condensation and explore the dynamics of the condensation and some of the relevant cosmological implications.
We consider the line element of the flat Friedmann-Robertson Walker metric
\begin{eqnarray}\label{eq2}
ds^{2}=-c^{2}dt^{2}+a^{2}(t)(dx^{2}+dy^{2}+dz^{2}),
\end{eqnarray}
where $a$ is the scale factor presenting the cosmological expansion. We describe the matter   by the perfect fluid energy momentum tensor,
\begin{eqnarray}\label{eq3}
T^{\mu\nu}=(\rho +p)u^{\mu}u^{\nu}+pg^{\mu\nu}.
\end{eqnarray}
where ($\rho$, $p$) stands for  ($\rho_{b}$, $p_{b}=0$),  ($\rho_{{\rm rad}}$, $p_{{\rm rad}}$), and
($\rho_{\chi}$, $p_{\chi}$) for baryonic matter, radiation and dark matter,
respectively. Note that we do not consider the interaction between these energy terms, in other words the energy of these components are individually satisfying the conservation equation.\\\\
Taking into account the well-known loop quantum gravity constraints and coupling the matter to classical phase space of FLRW universe, we obtain \cite{17}
\begin{eqnarray}\label{eq4}
\dot{V}=\frac{3V}{\lambda \gamma}\sin(\lambda b)\cos(\lambda b),
\end{eqnarray}
\begin{eqnarray}\label{eq5}
{\cal H}_{matt}=\frac{3}{4\gamma \lambda^{2}}V\sin^{2}(\lambda b),
\end{eqnarray}
where ${\cal H}_{matt}$ is the matter Hamiltonian, $\gamma\approx0.24 $\cite{18} represents the Barbero-Immirzi parameter and also $\lambda\approx2.27$ is the length gap. Meanwhile, $b$ is the conjugate momentum and $V$ is proportional to the physical volume of a cubical cell, with unit comoving volume $V=\frac{a^{3}}{2\pi \gamma}$. It is noticeable that dot means derivative with respect to the time.
By rewriting the equations (\ref{eq4}) and (\ref{eq5}), we have
\begin{eqnarray}\label{eq6}
H=\frac{\sin(2\lambda b)}{2\gamma \lambda},
\end{eqnarray}
\begin{eqnarray}\label{eq7}
\frac{\sin^{2}(\lambda b)}{\gamma^{2}\lambda^{2}}=\frac{8\pi}{3}\rho,
\end{eqnarray}
where $H$ is the Hubble parameter and the matter density is limited as $\rho<\rho_{c}$ where $\rho_{c}=\frac{3}{8\pi \gamma^{2}\lambda^{2}}$. Combination of  (\ref{eq6}) and  (\ref{eq7}) reduces to the modified FLRW equation in loop quantum cosmology model \cite{19}
\begin{eqnarray}\label{eq8}
3\frac{\dot{a}^{2}}{a^{2}}=8\pi G\rho\left(1-\frac{\rho}{\rho_{c}}\right).
\end{eqnarray}
As we mentioned, the effective equations have been derived for
FLRW loop quantum cosmologies considering different matter sectors at early universe \cite{multiple}. So, we may use the modified FLRW equation in loop quantum cosmology model, namely Eq.(\ref{eq8}), and assume the matter density $\rho$ as the combination of different matter components $\rho_{b}$, $\rho_{{\rm rad}}$, $\rho_{\chi}$, and $\Lambda$ indicating the baryonic
matter,
radiation, dark matter and cosmological constant respectively, so that
\begin{eqnarray}\label{eq8'}
3\frac{\dot{a}^{2}}{a^{2}}=8\pi G\left(\rho_{b}+\rho_{{\rm rad}}+\rho_{\chi}+\Lambda\right)\left(1-\frac{(\rho_{b}+\rho_{{\rm rad}}+\rho_{\chi}+\Lambda)}{\rho_{c}}\right).
\end{eqnarray}
 The second modified FLRW equation is also given by
\begin{eqnarray}\label{eq9}
2\frac{\ddot{a}}{a}+\frac{\dot{a}^{2}}{a^{2}}=8\pi G\left(\frac{\left(\rho_{b}+\rho_{{\rm rad}}+\rho_{\chi}\right)^{2}-\Lambda^{2}}{\rho_{c}}+\Lambda-\frac{\left(p_{\rm b}+p_{\rm rad}+p_{\Lambda}\right)}{c^{2}}\left(1-2\frac{(\rho_{b}+\rho_{{\rm rad}}+\rho_{\chi}+\Lambda)}{\rho_{c}}\right)\right),
\end{eqnarray}
 and also for the energy density conservation equation  we have
\begin{eqnarray}\label{eq10}
\dot{\rho_{i}}+3(\rho_{i}+\frac{p_{i}}{c^{2}})\frac{\dot{a}}{a}=0\:,\:\:(i\equiv
b,{ rad},\chi).
\end{eqnarray}

Now, regarding the cosmological evolution of radiation and baryonic matter we consider the relations $\rho_{{\rm rad}}=\rho_{{\rm rad,0}}/(a/a_{0})^{4}$ and $\rho_{b}=\rho_{b,0}/(a/a_{0})^{3}$, respectively; meanwhile $\rho_{b,0}$ and $\rho_{{\rm rad,0}}$ are the energy densities corresponding to their values at $a=a_{0}$, and also for the dark matter we work with a general form of density as $\rho_{\chi}=\rho_{\chi,0}/f(a/a_{0})$. It should be noted that $f(a/a_{0})$ is an arbitrary function which depends on the special dark matter model. As we know we have the critical density as $\rho_{{\rm cr,0}}=3H^{2}_{0}$, where $H_{0}$ is the value of the Hubble parameter at $a=a_{0}$. In addition, we can write dimensionless parameters known as density parameters as $\Omega_{i,0}=\rho_{i,0}/\rho_{{\rm cr,0}}$, with $i=b,{ rad},\chi$. Using these relations helps us to write the following new form of modified FLRW equation
\begin{eqnarray}\label{eq11}
\left(\frac{\dot{a}}{a}\right)^{2}=H^{2}_{0}\left(\frac{\Omega_{b,0}}{(\frac{a}{a_{0}})^{3}}+\frac{\Omega_{{\rm rad,0}}}{(\frac{a}{a_{0}})^{4}}+\frac{\rho_{\chi,0}}{f(\frac{a}{a_{0}})}+\Omega_{\Lambda}\right)-\frac{H^{4}_{0}}{\rho_{c}}\left(\frac{\Omega_{b,0}}{(\frac{a}{a_{0}})^{3}}+\frac{\Omega_{{\rm rad,0}}}{(\frac{a}{a_{0}})^{4}}+\frac{\rho_{\chi,0}}{f(\frac{a}{a_{0}})}+\Omega_{\Lambda}\right)^{2},
\end{eqnarray}
where $\Omega_{\Lambda}$ is the dark energy density parameter. Thus, we can write the constraint
$$
\Omega_{b,0}+\Omega_{{\rm rad,0}}+\Omega_{\chi,0}+\Omega_{\Lambda}=1.
$$

Now, we suppose that at early universe the  bosonic particles, with mass $m_{\chi}$
and a very high
temperature $T$, were in equilibrium state
with other particles in a hot relativistic plasma. After the expansion of universe and rapid decrease
of the temperature it was decoupled from this equilibrium state at a high
chemical potential $\mu \gg m_{\chi}$ or low decoupling temperature $T_{_{D}}\ll T$. At this decoupling temperature, we  assume that the
bosonic particles are still in kinetic equilibrium among themselves in an isotropic gas with
low temperature. Consequently, for the spatial number density we have
\begin{eqnarray}\label{eq12}
n=\frac{4\pi g}{h^{3}}\int   p^{2} f(p) dp,
\end{eqnarray}
where $g$ is the helicity states number, $h$ is planck's constant, and $f(p)$ is defined as follows
\begin{eqnarray}\label{eq13}
f(p)=\frac{1}{\left(\exp[E-\mu]-1\right)},
\end{eqnarray}
where $E$ is the energy $E=\sqrt{p^{2}+m_{\chi}^{2}c^{4}}$ and $p$ is the  momentum of the particle. Due to the red-shift in the momentum of particles, we
have $p(t)a(t)=p_{_{D}}a_{_{D}}$ where $p_{_{D}}$ and $a_{_{D}}$ denote for the momentum of particles and the scale factor of universe, respectively,  at the  temperature $T_{_{D}}$ denoting for the decoupling temperature from the rest of plasma. Besides, we have the scaling evolution relation for the number density given by $n_{\chi}\sim a^{-3}$ \cite{20}. For the extreme-relativistic
case $E\approx pc$ ( $\mu=\mu_{_{D}}a_{_{D}}/a$ and $T=T_{_{D}}a_{_{D}}/a$), the distribution function is derived as $f_{ER}(p)=\left(\exp[(pc-\mu)]-1\right)^{-1}$, hence for the nonrelativistic case $E-\mu\approx p^{2}/2m_{\chi}-\mu_{_{kin}}$ ( $\mu_{_{kin}}\equiv\mu-m_{\chi}c^{2}$ ) the distribution function is given by $f_{{\rm NR}}(p)=\left(\exp[(p^{2}/2m_{\chi}-\mu_{_{kin}})]-1\right)^{-1}$, where the evolution equations are as $\mu_{_{kin}}=\mu_{_{kin,D}}(a_{_{D}}/a)^{2}$ and $T=T_{_{D}}/(a_{_{D}}/a)^{2}$ \cite{20}. For  a frozen distribution of dark matter, the  associated energy density $\epsilon$ and  kinetic energy-momentum tensor $T^{\mu}_{\nu}$ are as follows, respectively
\begin{eqnarray}\label{eq14}
\epsilon=\frac{g}{3h^{3}}\int E f(p)\: d^{3}p,
\end{eqnarray}
\begin{eqnarray}\label{eq15}
T^{\mu}_{\nu}=\frac{g}{h^{3}}\int\frac{p^{\mu}p_{{\nu}}}{p^{0}}f(p)d^{3}p.
\end{eqnarray}
The mentioned pressure is described by
\begin{eqnarray}\label{eq16}
p=\frac{g}{3h^{3}}\int p \emph{v} f(p) d^{3}p =\frac{gc^{2}}{3h^{3}}\int\frac{p^{2}}{E}f(p)d^{3}p,
\end{eqnarray}
where we have used $\emph{v}=pc^{2}/E$ \cite{21}. The density of dark matter $\rho_{\chi}$ for the nonrelativistic case with $E=m_{\chi}c^{2}$ and $P\approx m_{\chi}\emph{v}_{\chi}$ is given by $\rho_{\chi}=m_{\chi}n_{\chi}$ whereas the dark matter pressure is given by
\begin{eqnarray}\label{eq17}
p_{\chi}=\frac{gc^{2}}{3h^{3}}\int\frac{p^{2}}{E}f(p)d^{3}p\approx\frac{4\pi
g}{3h^{3}}\int\frac{p^{4}}{m_{\chi}}dp,
\end{eqnarray}
 which leads to
\begin{eqnarray}\label{eq18}
p_{\chi}=c^{2}\sigma^{2}\rho_{\chi},
\end{eqnarray}
where we have used $\sigma^{2}=<\vec{\emph{v}_{x}}^{2}>/3c^{2}$,  $\sigma$ being the one-dimensional velocity dispersion. By means of the dark matter conservation equation
\begin{eqnarray}\label{eq19}
\dot{\rho_{\chi}}+3\frac{\dot{a}}{a}\rho_{\chi}(1+\sigma^{2})=0,
\end{eqnarray}
we can find the following general solution
\begin{eqnarray}\label{eq20}
\rho_{\chi}=\frac{\rho_{\chi,0}}{\left(a/a_{0}\right)^{3(1+\sigma^{2})}},
\end{eqnarray}
where $\rho_{\chi,0}$ is the density of dark matter  at the present value of  scale factor $a=a_{0}$. In the standard model, for the description of the dynamics of our universe with a normal dark matter we work with the following equation$$
\frac{\dot{a}^{2}}{a^{2}}=H^{2}_{0}\left[\frac{\Omega_{b,0}}{\left(a/a_{0}\right)^{3}}+\frac{\Omega_{{\rm rad,0}}}{\left(a/a_{0}\right)^{4}}+\frac{\Omega_{\chi,0}}{\left(a/a_{0}\right)^{3(1+\sigma^{2})}}+\Omega_{\Lambda}\right].
$$
However, in LQC model we will have the following modification
\begin{eqnarray}\label{eq21}
\frac{\dot{a}^{2}}{a^{2}}=
H_{0}^{2}\left(\frac{\Omega_{b,0}}{\left(a/a_{0}\right)^{3}}+\frac{\Omega_{{\rm rad,0}}}{\left(a/a_{0}\right)^{4}}+
\frac{\Omega_{\chi,0}}{\left(a/a_{0}\right)^{3(1+\sigma^{2})}}+\Omega_{\Lambda}\right)-\\\nonumber
\frac{H_{0}^{2}}{\rho_{c}}
\left(\frac{\Omega_{b,0}}{\left(a/a_{0}\right)^{3}}+\frac{\Omega_{{\rm rad,0}}}{\left(a/a_{0}\right)^{4}}+
\frac{\Omega_{\chi,0}}{\left(a/a_{0}\right)^{3(1+\sigma^{2})}}+\Omega_{\Lambda}\right)^{2}.
\end{eqnarray}
Since we aim to extract numerical results for dynamical BEC quantities, we  have to mention the values of several parameters and quantities such as the Hubble constant $H_{0}=70$ $km/s/Mpc=2.27\times10^{-18}$ $s^{-1}$,  the Hubble time $t_{H}={H_{0}^{-1}}=4.39\times10^{17}$ $s$, the critical density $\rho_{{\rm cr,0}}=9.24\times10^{-30}$ $gr/cm^{3}$, and the present day dark matter densities $\Omega_{b,0}=0.045$, $\Omega_{\chi,0}\approx0.228$, $\Omega_{{\rm rad,0}}=8.24\times10^{-5}$, $\Omega_{\Lambda}=0.73$ respectively
\cite{22}. It is worth to remind that the following dark matter case is non-relativistic so the global cosmological evolution of the universe will not be affected by the variation of the numerical values of $\sigma^{2}$.

As is explained in the introduction, in the very low temperature all dilute Bose gas particles are condensed  to the same quantum ground state which
leads to the BEC formation. When the particle's wavelengths overlap, they will experience correlation with each other and the mean inter-particles distance $l$ becomes definitely smaller than the thermal wavelength $\lambda_{T}$. This is possible when the  temperature is almost $T_{{\rm cr}}\approx2\pi\hbar^{2}\rho^{2/3}/m^{5/3}$ $k_{B}$, where $m$ is the particle mass in the condensate, $k_{{\rm B}}$ is Boltzmann's constant \cite{4} and $\rho$ is the density. Under the condition that the universe is sufficiently large and temperature is low enough, there will be a coherent state in progress due to the overlap of the particles
wavelengths. Our main
assumption is that the dark matter halos may be constructed by a strongly coupled dilute BEC at absolute zero. As a result, all the particles are in the condensate. Therefore in this situation, just binary collisions at low energy are considerable. These collisions are specified by just one parameter
$l_{a}$, {\it s-wave scattering length},  which is thoroughly independent of the two-body potential. Consequently, the interaction potential can be replaced by an effective interaction $V_{l}(\vec{r}'-\vec{r})=\lambda ~\delta(\vec{r}'-\vec{r})$, where $\lambda=4\pi\hbar^{2}l_{a}/m_{\chi}$ \cite{4}. The Gross-Pitaevskii (GP) equation, describing the ground state properties of the dark matter, for the dark matter halos can be obtained from the GP energy functional as follows
\begin{eqnarray}\label{eq22}
E[\Psi]=\int\left[\frac{\hbar^{2}}{2m_{\chi}}|\nabla\Psi(\vec{r})|^{2}+\frac{U_{0}}{2}|\Psi(\vec{r})|^{4}\right]d\vec{r}-
\frac{1}{2}Gm^{2}_{\chi}\int\int\frac{|\Psi(\vec{r})|^{2}|\Psi(\vec{r}^{'})|^{2}}{|\vec{r}-\vec{r}'|}d\vec{r}d\vec{r}^{'},
\end{eqnarray}
where $U_{0}=4\pi\hbar^{2}l_{a}/m_{\chi}$ \cite{4} and obviously we have applied $\Psi(\vec{r})$ as the wave function of the condensate. In the equation (\ref{eq22}), the first and second terms are the quantum pressure and the interaction energy, respectively, and also the third term is the gravitational potential energy. It should be noted that the condensate dark matter mass density is as follows
\begin{eqnarray}\label{eq23}
\rho_{\chi}(\vec{r})=m_{\chi}~|\Psi(\vec{r})|^{2}=m_{\chi}~\rho(\vec{r},t),
\end{eqnarray}
and that the wave function satisfies  the normalization condition  $N=\int|\Psi(\vec{r})|^{2}d\vec{r}$, where $N$ is the total number of particles. Eventually, by imposing the variational procedure $\delta E[\Psi]-\mu~\delta\int|\Psi(\vec{r})|^{2}d\vec{r}=0$, the GP equation is extracted
\begin{eqnarray}\label{eq24}
-\frac{\hbar^{2}}{2m_{\chi}}\nabla^{2}\Psi(\vec{r})+m_{\chi}V(\vec{r})~\Psi(\vec{r})+U_{0}|\Psi(\vec{r})|^{2}~\Psi(\vec{r})=\mu~\Psi(\vec{r}),
\end{eqnarray}
while $\mu$ plays the role of chemical potential. In addition the gravitational potential $V$ follows the well-known poison equation $\nabla^{2}V=4\pi G\rho$. Having considered time dependency, the generalized GP equation, showing a gravitationally trapped rotating BEC, is found as
\begin{eqnarray}\label{eq25}
i \hbar \frac {\partial}{\partial t}\Psi(\vec{r},t)=\left[-\frac{\hbar^{2}}{2m_{\chi}}\nabla^{2}+m_{\chi}
 V(\vec{r})+U_{0} |\Psi(\vec{r},t)|^{2}\right]\Psi(\vec{r},t).
\end{eqnarray}

By means of the well-known Madelung representation of the wave function, we can reach to our purpose more easily than using the last generalized GP equation. In order to go trough this, we should consider the following representation of $\Psi$
\begin{eqnarray}\label{eq26}
\Psi(\vec{r},t)=\sqrt{\rho(\vec{r},t)}~\exp \left[\frac{i}{\hbar}S(\vec{r},t)\right],
\end{eqnarray}
where  $S(\vec{r},t)$ represents the classical action. Inserting the last expression of $\Psi(\vec{r},t)$ into equation (\ref{eq25}) leads to the following differential equations
\begin{eqnarray}\label{eq27}
\frac{\partial \rho_{\chi}}{\partial t} + \nabla\cdot(\rho_{\chi}\vec{\emph{v}})=0,
\end{eqnarray}
\begin{eqnarray}\label{eq28}
\rho_{\chi} \left[\frac{\partial\vec{\emph{v}}}{\partial t} + (\vec{\emph{v}}\cdot\nabla)
\vec{\emph{v}}\right]=- \nabla p_{\chi}\left(\frac{\rho_{\chi}}{m_{\chi}}\right)
- \rho_{\chi}\nabla\left(\frac{V}{m_{\chi}}\right)-\nabla V_{Q},
\end{eqnarray}
where $\vec{\emph{v}}=\nabla S/m_{\chi}$ is the velocity of the quantum fluid and  $V_{Q}$ is the quantum potential, given by
\begin{eqnarray}\label{eq29}
V_{Q} = -(\frac{\hbar^{2}}{2m_{\chi}})\frac{ \nabla^{2 }\sqrt{\rho\chi}}{\sqrt{\rho\chi}}.
\end{eqnarray}
By defining $u_{0}=\frac{2\pi\hbar^{2}l_{a}}{m_{\chi}^{3}}$, the condensate effective pressure is obtained
\begin{eqnarray}\label{eq30}
p_{\chi} \left(\frac{\rho_{\chi}}{m_{\chi}}\right) = u_{0} \rho^{2}_{\chi}.
\end{eqnarray}

The quantum pressure term makes a great contribution just close to the boundary of the condensate provided that the number of particles in the gravitationally bounded BEC becomes large enough. As a result, the quantum stress term in the condensate equation of motion can be ignored, which is so called Thomas-Fermi approximation. When the number of particles in the condensate goes to infinity, the mentioned Thomas-Fermi approximation becomes accurate. It is also related to the classical limit of the theory. According to its definition, the velocity field is irrotational following the condition $\nabla\times\vec{\emph{v}}=0$.
In accordance with equations (\ref{eq10}) and (\ref{eq30}) we can write
\begin{eqnarray}\label{eq31}
\dot{\rho_{\chi}}+3\rho_{\chi} \frac{\dot{a}}{a}\left(1+\frac{u_{0}}{c^{2}}\rho_{\chi}\right)=0,
\end{eqnarray}
which has the following general solution with the integration constant $C_{\chi}$
\begin{eqnarray}\label{eq32}
\rho_{\chi} = \frac{C_{\chi }}{\left(a/a_{0}\right)^{3} - \left(u_{0}/c^{2}\right)
C_{\chi}}.
\end{eqnarray}
Having imposed the condition $\rho_{\chi}=\rho_{\chi,0}$ for $a=a_{0}$, we can find the density of the condensate as
\begin{eqnarray}\label{eq33}
\rho_{\chi} = \frac{c^{2}}{u_{0}} \frac{\rho_{0\chi}}{(a/a_{0})^{3} - \rho_{0\chi}},
\end{eqnarray}
where
\begin{eqnarray}\label{eq34}
\rho_{0\chi}=\frac{\rho_{\chi,0}u_{0}c^{2}}{1+\rho_{\chi,0}u_{0}/c^{2}}.
\end{eqnarray}

\section{Loop quantum cosmological dynamics of the Bose-Einstein condensation}

For an interacting Bose system case, where all dark matter particles are
supposed to be  in a single-particle state, the transition procedure from the normal to the condensed phase, and the order of the phase transition has been extensively studied in the recent literature \cite{14}-\cite{16}. In principle, BEC represents a spontaneous breaking of $U(1)$ gauge symmetry, with the condensate fraction showing the order parameter, leading to the second order phase transition \cite{23}. Nevertheless, it has been shown that BEC is a first-order phase transition\cite{24}, \cite{25}. From another point of view, in thermodynamics
the first-order phase transitions result in a  genuine mathematical singularity. The question that ``can finite systems in nature  explicitly show such a behavior?" is a mysterious controversial question in physics \cite{26}. Nonetheless, a significant study about the thermodynamic instability in an ideal Bose gas, confined in a cubic box, has shown that a system including a finite number of particles can exhibit an interrupted phase transition, describing a real mathematical singularity, with constant pressure \cite{25}. Of course the mentioned result was obtained without considering a continuous approximation or a thermodynamic limit. Therefore, according to above discussion, we will assume that the intrinsic dynamics of BEC is described in the framework of a first-order phase transition.

{\subsection{Cosmological parameters at the condensation point}}

As we know, in general, the chemical potential $\mu$ in a physical system is a function which depends on the temperature and particle density $n=N/V$ ($n$ is the total particle number and $V$ is the volume). By considering the property of extensivity of the Helmholtz free energy $F=F(N,V,T)$, we are allowed to write not only $F=Vf(n,T)$ but also $F=N\tilde{f}(\emph{v},T)$, where $\emph{v}=V/N=n^{-1}$ and $f=n\tilde{f}$. Using the Helmholtz free energy we can obtain $\mu(n,T)=\left(\partial f/\partial n\right)_{T}$ and $p(n,T)=-(\partial \tilde{f}/\partial \emph{v})_{T}$ having the same information.

Imposing the thermodynamical requirements, the chemical potential and the pressure are supposed to be single-valued which means that for arbitrary values of $n$, $\emph{v}$ and $T$, there must be just one choice for $\mu$ or $p$, respectively \cite{24}. Hence, the pressure at the transition moment during the BEC process must be continuous. Imposing this thermodynamic condition on equations (\ref{eq18}) and (\ref{eq30}), helps us to fix the transition density $\rho^{{\rm cr}}_{\chi}$ as follows
\begin{eqnarray}\label{eq35}
\rho^{{\rm cr}}_{\chi} = \frac{c^{2}\sigma^{2}}{u_{0}} = \frac{c^{2}\sigma^{2}m_{\chi}^{3}}{2\pi \hbar^{2}l_{a}}.
\end{eqnarray}

Apparently, in order to have the numerical value of the transition density, we should know the scattering length, the dark matter particle mass and the particle velocity dispersion. Supposing typically the mass of dark matter particle, the mean velocity square, and the scattering length to be of the order of $1$ $eV$,  $10^{15}$ $cm^{2}/s^{2}$ and  $10^{-10}$ $cm$, respectively, we can write
\begin{eqnarray}\label{eq36}
\rho^{{\rm cr}}_{\chi} = 3.868 \times 10^{-21}\times \left(\frac{\sigma^{2}}{3\times10^{-6}}\right)
\times \left(\frac{m_{\chi}}{10^{-33}g}\right)^{3}\times\left(\frac{l_{a}}{10^{-10}cm}\right)^{-1}~g/cm^{3}.
\end{eqnarray}
Therefore, the critical temperature at the BEC moment is obtained by
\begin{eqnarray}\label{eq37}
T_{\rm cr} \approx \frac{2 \pi \hbar^{2}}{\zeta(3/2)^{2/3} m_{\chi}^{5/3}k_{{\rm B}}} (\rho^{{\rm cr}}_{\chi})^{2/3} = \frac{(2\pi\hbar^{2})^{1/3}c^{4/3}}{ \zeta(3/2)^{2/3} k_{{\rm B}}} \frac{(\sigma^{2})^{2/3} m_{\chi}^{1/3}}{l_{a}^{2/3}},
\end{eqnarray}
where $\zeta (3/2)$ is the Riemann zeta function, or equivalently
\begin{eqnarray}\label{eq38}
T_{{\rm cr}} \approx 6.57 \times 10^{3} \times \left(\frac{m_{\chi}}{10^{-33}g}\right)^{1/3}
 \times\left(\frac{\sigma^{2}}{3\times10^{-6}}\right)^{2/3} \times \left(\frac{l_{a}}{10^{-10}cm}\right)^{-2/3}~K.
\end{eqnarray}
The critical pressure of the dark matter fluid at the critical point is given by

\begin{eqnarray}\label{eq39}
p_{{\rm cr}}=1.04\times10^{-5} \times \left(\frac{\sigma^{2}}{3\times10^{-6}}\right)^{2}
\times \left(\frac{m_{\chi}}{10^{-33}g}\right)^{3} \times \left(\frac{l_{a}}{10^{-10}cm}\right)^{-1}~dyne/cm^{2}.
\end{eqnarray}
Using the equation (\ref{eq35}) we can calculate the critical value of scale factor as
\begin{eqnarray}\label{eq40}
\frac{a_{{\rm cr}}}{a_{0}} = \left(\frac{\rho_{\chi,0}u_{0}}{c^{2}\sigma^{2}}\right)^{1/3(1+\sigma^{2})}
 = \left(\frac{2 \pi \hbar^{2} l_{a} \rho_{{\rm cr,0}} \Omega_{\chi,0}}{c^{2}
 \sigma^{2} m_{\chi}^{3}}\right)^{1/3 ( 1 + \sigma^{2} )}.
\end{eqnarray}
Consequently, we can find the critical red-shift as follows

\begin{eqnarray}\label{eq41}
1 + z_{{\rm cr}} = \left(\frac{2 \pi \hbar^{2} l_{a} \rho_{{\rm cr,0}} \Omega_{\chi,0}}{c^{2}
\sigma^{2} m_{\chi}^{3}}\right)^{-1/3( 1 + \sigma^{2} )}.
\end{eqnarray}
By inserting the above mentioned numerical values of the parameters in the two last relations we have

 \begin{eqnarray}\label{eq42}
\frac{a_{{\rm cr}}}{a_{0}} = 8.17 \times 10^{-4} \times \left(\frac{m_{\chi}}{10^{-33}g}\right)^{-(1
+\sigma^{2})} \times \left(\frac{\sigma^{2}}{3 \times10^{-6}}\right)^{-1/3( 1 + \sigma^{2})}\times
\left(\frac{l_{a}}{10^{-10}cm}\right)^{1/3(1+\sigma^{2})},
\end{eqnarray}
and
\begin{eqnarray}\label{eq43}
1 +z_{{\rm cr}} = 1.22 \times 10^{3} \times \left(\frac{m_{\chi}}{10^{-33}
g}\right)^{( 1 + \sigma^{2})} \times \left(\frac{\sigma^{2}}{3 \times 10^{-6}}\right).
\end{eqnarray}
The obtained critical values of density, temperature, pressure,   scale factor and  redshift are important in the study of subsequent (LQC) evolution of the the universe during the Bose-Einstein condensation
toward  the end time of  phase transition
when  the universe enters in the complete Bose-Einstein condensed dark matter phase.
We will use these critical values in the study of this evolution in the following
subsection. 
\subsection{Loop quantum cosmological evolution during the Bose-Einstein condensation phase}

In the first-order phase transition, we have constant temperature and pressure,
respectively as $T=T_{{\rm cr}}$ and $p=p_{{\rm cr}}$.
Moreover,    the enthalpy $W=(\rho+p)a^{3}$ and entropy $S=sa^{3}$ are also
conserved quantities. The phase transition begins through a decrease in the density of dark matter and increase in the density of Bose-Einstein
condensed state. This is because the density $\rho^{{\rm cr}}_{\chi}(T_{{\rm cr}})\equiv\rho^{nor}_{\chi}$ is converted to $\rho_{\chi}(T_{{\rm cr}})\equiv\rho^{{\rm BEC}}_{\chi}$ during the phase transition. One may define a  useful time dependent parameter $h(t)$, playing the role of $\rho_{\chi}(t)$, which is called the volume fraction of the matter in the Bose-Einstein condensed phase
\begin{eqnarray}\label{eq44}
h(t)=\frac{\rho_{\chi}(t)-\rho^{{\rm nor}}_{\chi}}{\rho^{{\rm BEC}}_{\chi}-\rho^{{\rm nor}}_{\chi}}.
\end{eqnarray}
By simplifying (\ref{eq44}), we can extract the time evolving dark matter energy density during the transition process
as
\begin{eqnarray}\label{eq45}
\rho_{\chi}(t) = \rho^{{\rm BEC}}_{\chi} h(t) + \rho^{{\rm nor}}_{\chi}[1
- h(t)] = \rho^{{\rm nor}}_{\chi}[1 + n_{\chi} h(t)],
\end{eqnarray}
where we have defined
\begin{eqnarray}\label{eq46}
n_{\chi} = \frac{\rho^{\rm BEC}_{\chi} - \rho^{\rm nor}_{\chi}}{\rho^{\rm nor}_{\chi}}.
\end{eqnarray}
By looking at equation(\ref{eq44}), we can determine the beginning and final points as  $h(t_{{\rm cr}})=0$ ($t_{{\rm cr}}$ being the beginning time of transition in which $ \rho_{\chi}(t_{{\rm cr}}) \equiv \rho^{\rm nor}_{\chi}$)
and $h (t_{\rm BEC})=1$ ($t_{{\rm BEC}}$ being the end time of transition in which $\rho_{\chi} (t_{\rm BEC}) \equiv \rho^{\rm BEC}_{\chi}$), respectively. At the end time of  phase transition  the universe enters in the Bose-Einstein condensed dark matter phase.
As mentioned, by using $h(t)$ instead of $\rho_{\chi}(t)$ in equation (\ref{eq10}), we can find the following relation
\begin{eqnarray}\label{eq47}
&&\frac{\dot{a}}{a} = - \frac{1}{3} \frac{\left(\rho^{\rm BEC}_{\chi}-\rho^{\rm nor}_{\chi}\right)\dot{h}}{\rho^{\rm nor}_{\chi}+p_{\rm cr}+\left(\rho^{\rm BEC}_{\chi}-\rho^{\rm nor}_{\chi} \right)} = -\frac{1}{3}\frac{r\dot{h}}{ 1 + rh},
\end{eqnarray}
where
\begin{eqnarray}\label{eq48}
r = \frac{\rho^{\rm BEC}_{\chi}-\rho^{\rm nor}_{\chi}}{\rho^{\rm nor}_{\chi}+
p_{\rm cr}} = \frac{n_{\chi}}{1 + p_{\rm cr}/\rho^{\rm nor}_{\chi}}.
\end{eqnarray}
The last expression is implying that in general we have $r<0$, $r\in(-1,0)$ and also $n_{\chi}<0$, because $\rho^{{\rm BEC}}_{\chi}<\rho^{{\rm nor}}_{\chi}$. Clearly, imposing the condition $h(t_{{\rm cr}})=0$,
we can obtain the relation between the scale factor and $h(t)$ from equation (\ref{eq47}) as
\begin{eqnarray}\label{eq49}
a(t)= a_{{\rm cr}}[ 1 + r h(t)]^{-1/3},~~~~    t\in(t_{{\rm cr}},t_{{\rm BEC}}),
\end{eqnarray}
where  $a_{{\rm cr}}=a(t_{{\rm cr}})$. The value of scale factor at the end of  phase transition is as follows
\begin{eqnarray}\label{eq50}
a_{{\rm BEC}} = a(t_{{\rm BEC}}) = a_{{\rm cr}}(1 + r)^{-1/3}.
\end{eqnarray}
This is an important result, since this condensation process has modified the expansion rate of the universe. Subsequently, we can find the evolution equations for the baryonic matter and the radiation density during the phase transition, respectively as
\begin{eqnarray}\label{eq51}
\rho_{b}=\frac{\rho_{b,0}}{(a_{{\rm cr}}/a_{0})^{3}}[ 1 + r h(t)], t\in(t_{{\rm cr,t_{{\rm BEC}}}}),
\end{eqnarray}
and
\begin{eqnarray}\label{eq52}
\rho_{{\rm rad}}=\frac{\rho_{{\rm rad,0}}}{(a_{{\rm cr}}/a_{0})^{4}}[1 +
r h(t)]^{\frac{4}{3}},  t \in (t_{{\rm cr, t_{\rm BEC}}}).
\end{eqnarray}

 Using all these equations, we consider the time evolution of the volume fraction of the condensed matter dark energy density during the BEC process, which is describing the dynamics of this process in the cosmic history in LQC model, as
\begin{eqnarray}\label{eq53}
&&\frac{dh}{d\tau}=-3(\frac{1}{r}+h)\sqrt{\frac{\Omega_{b,0}}{\left(\frac{a_{{\rm cr}}}{a_{0}}\right)^{3}}(1+rh) +\frac{\Omega_{{\rm rad,0}}}{\left(\frac{a_{{\rm cr}}}{a_{0}}\right)^{4}}(1+rh)^{4/3}+\Omega_{\chi,{\rm nor}}(1+n_{\chi}h)+\Omega_{\Lambda}}\times\\\nonumber&&\sqrt{1
-\frac{1}{\rho_{c}}\left(\frac{\Omega_{b,0}}{\left(\frac{a_{{\rm cr}}}{a_{0}}\right)^{3}}(1+rh) +\frac{\Omega_{{\rm rad,0}}}{\left(\frac{a_{{\rm cr}}}{a_{0}}\right)^{4}}(1+rh)^{4/3}+\Omega_{\chi,{\rm nor}}(1+n_{\chi}h)+\Omega_{\Lambda}\right)}.
\end{eqnarray}

\
Here, we have used $\tau$ as a dimensionless time variable $\tau=H_{0}t$
and  $\Omega_{\chi,{\rm nor}}=\rho^{{\rm nor}}_{\chi}/\rho_{{\rm cr,0}}$. Considering the equation (\ref{eq48}) and   $p_{{\rm cr}}/\rho^{{\rm nor}}_{\chi}=\sigma^{2}\ll1$, we  deduce the approximation $r\approx n_{\chi}$ in the following calculations. Another approximation comes from neglecting the radiation energy density contribution, so the equation (\ref{eq53}) can be integrated to extract $h(t)$
\begin{eqnarray}\label{eq54}
&&h(t)=\frac{\Omega_{\Lambda}^{2}}{r\Omega_{\rm tr}}\times\nonumber\\&&\left(\frac{\rho_{c}^{2}}{4\Omega_{\Lambda}^{4}}\left(\Omega_{\rm tr}+\Omega_{\Lambda}\right)\left(\Omega_{\rm tr}+\Omega_{\Lambda}-\rho_{c}\right)+\frac{\left(-1+M\right)^{2}}{4\Omega_{\Lambda}^{4}\left(\rho_{c}-\Omega_{\Lambda}\right)}\left(\Omega_{\rm tr}\rho_{c}-2\left(\Omega_{\rm tr}-\rho_{c}\right)\Omega_{\Lambda}-2\Omega_{\Lambda}^{2}\right)^{2}+\right.\nonumber\\&&\left.\frac{\left(-1+M\right)M}{\Omega_{\Lambda}^{4}\sqrt{\Omega_{\Lambda}\left(1-\frac{\Omega_{\Lambda}}{\rho_{c}}\right)}}\times\sqrt{-\left(\Omega_{\Lambda}+\Omega_{\rm tr}\right)\left(\Omega_{\Lambda}+\Omega_{\rm tr}-\rho_{c}\right)\left(\Omega_{\rm tr}\rho_{c}-2\left(\Omega_{\rm tr}-\rho_{c}\right)\Omega_{\Lambda}-2\Omega_{\Lambda}^{2}\right)}\right)^{\frac{1}{2}}\nonumber
\\&&-\frac{\Omega_{\Lambda}+\Omega_{\rm tr}}{r\Omega_{\rm tr}}+\frac{\rho_{c}}{2r\Omega_{\rm tr}},
\end{eqnarray}
where we have defined
\begin{eqnarray}\label{eq55}
\Omega_{{\rm tr}}=\frac{\Omega_{b,0}}{\left(\frac{a_{{\rm cr}}}{a_{0}}\right)^{3}}+\Omega_{\chi,{\rm nor}},
\end{eqnarray}
and
\begin{eqnarray}\label{eq56}
M=e^{3H_{0}\left(t-t_{cr}\right)\sqrt{\Omega_{\Lambda}\left(1-\frac{\Omega_{\Lambda}}{\rho_{c}}\right)}}.
\end{eqnarray}
In order to obtain the necessary time interval for the entire converting process of
the total normal dark matter into the Bose-Einstein condensed phase, we put $h(t=t_{{\rm BEC}})=1$ in our last calculation
\begin{eqnarray}\label{eq57}
\Delta t_{{\rm con}}&=&t_{{\rm BEC}}-t_{{\rm cr}}
\\\nonumber&=&\frac{t_{H}}{3\Omega_{\Lambda}}\times\sqrt{\frac{\Omega_{\Lambda}}{1-\frac{\Omega_{\Lambda}}{\rho_{c}}}}\times\\\nonumber&\textrm{ln}&\left[\frac{2\sqrt{\Omega_{\Lambda}\left(1-\frac{\Omega_{\Lambda}}{\rho_{c}}\right)}\sqrt{\left(\frac{-\Omega_{\rm tr}^{2}}{\rho_{c}}+\Omega_{\rm tr}-2\frac{\Omega_{\rm tr}\Omega_{\Lambda}}{\rho_{c}}\right)(1+r)+\Omega_{\Lambda}\left(1-\frac{\Omega_{\Lambda}}{\rho_{c}}\right)}+2\Omega_{\Lambda}\left(1-\frac{\Omega_{\Lambda}\Omega_{\rm tr}}{\rho_{c}}\right)+\Omega_{\rm tr}}{\left(1+r\right)\left(2\sqrt{\Omega_{\Lambda}\left(1-\frac{\Omega_{\Lambda}}{\rho_{c}}\right)}\sqrt{\frac{-\Omega_{\rm tr}^{2}}{\rho_{c}}+\Omega_{\rm tr}-2\frac{\Omega_{\rm tr}\Omega_{\Lambda}}{\rho_{c}}+\Omega_{\Lambda}\left(1-\frac{\Omega_{\Lambda}}{\rho_{c}}\right)+2\Omega_{\Lambda}\left(1-\frac{\Omega_{\Lambda}+\Omega_{\rm tr}}{\rho_{c}}\right)+\Omega_{\rm tr}}\right)}\right].
\end{eqnarray}
By inserting the value of dark energy density parameter $\Omega_{\Lambda}$ as {0.68} and also considering the standard values $\rho_{c}\simeq0.82M_{p}^{4}\simeq1.84\times10^{170}$ ${\rm eV}/{\rm cm}^{3}$, $\ell_{a}=10^{-10}$ ${\rm cm}$, $m=1$ $ {\rm eV}$, and $\Omega_{{\rm tr}}=5.02\times10^{8}$ , we can obtain the following expression for the conversion time of the dark matter to the Bose-Einstein condensed phase
\begin{eqnarray}\label{eq58}
\Delta t_{{\rm cond}}=0.39\times\textrm{ln}\left[\frac{0.99+1.135\times10^{-9}\sqrt{5.02\times10^{8}+5.02\times10^{8}r}}{1+r}\right]\times t_{H}\rm s.
\end{eqnarray}
For instance,   for $r=-0.2$ and $t_{H}=4.39\times10^{17}\rm s$, we will have $\Delta t_{{\rm cond}}\simeq 3.82\times10^{16}\rm s$.
 Since it is thoroughly useful to depict the time evolution of $h(t)$, we have represented (\ref{eq54}) for different values of $r$ in Fig.1.

\subsection{The post condensation phase in LQC model}

After the complete conversion to the BEC phase, the post-condensation phase begins, which starts at $t=t_{{\rm BEC}}$ and $a_{{\rm BEC}}=a_{{\rm cr}}(1+r)^{-1/3}$.  Similar to the era, during the condensation
phase, 
in which the cosmological dynamics is affected by the BEC (see (\ref{eq50})), the Bose-Einstein condensed dark matter, in the post condensation phase, may change the cosmological dynamics of the universe in LQC model. Moreover,
the Bose-Einstein condensed dark matter may have considerable impact on the
subsequent structure formation in the universe. Therefore, the post condensation phase and the corresponding cosmological values in LQC model are important
and deserve more
scrutiny.

According to (\ref{eq33}), we can obtain the cosmological density of the dark matter as follows
\begin{eqnarray}\label{eq59}
\rho^{{\rm BEC}}_{\chi} = \frac{c^{2}}{u_{0}} \frac{\rho_{0\chi}( 1 + r )}{(a_{{\rm cr}}/a_{0})^{3}- \rho_{0\chi}( 1 + r )},
\end{eqnarray}
where, the numerical value of $\rho_{0\chi}$ is obtained from (\ref{eq34})

\begin{eqnarray}\label{eq60}
\rho_{0\chi} = \frac{1.63 \times 10^{-15} \times (l_{a}/10^{-10}cm)(m/10^{-33}gr)^{-3}}{1
+ 1.63 \times 10^{-15} \times (\l_{a}/10^{-10}cm)(m/10^{-33}gr)^{-3}}.
\end{eqnarray}
Clearly, the density (\ref{eq59})  should be positive which leads to the condition $(a_{{\rm cr}}/a_{0})>_{\rho0\chi^{1/3}}(1+r)^{1/3}$.
Simplifying the equation (\ref{eq21}) and defining $\Omega_{{\rm BE}}=\frac{\Omega_{\chi,0}}{1+\rho_{\chi,0}u_{0}/c^{2}}$ gives us the time evolution of the scale factor in the BEC phase as follows

\begin{eqnarray}\label{eq61}
\frac{d\left(a/a_{0}\right)}{dt}=H_{0}\sqrt{\Omega_{BE}}\frac{a/a_{0}}{\sqrt{\left(a/a_{0}\right)^{3}-\rho_{0\chi}}}\sqrt{1-\frac{H_{0}^{2}}{\rho_{c}}\frac{\Omega_{BE}}{\left(a/a_{0}\right)^{3}-\rho_{0\chi}}}.
\end{eqnarray}
By integration of this relation, we obtain exactly an expression with the following form
\begin{eqnarray}\label{eq62}
\sqrt{\Omega_{{\rm BE}}}H_{0}(t-C)&=&\frac{2}{3} \sqrt{(a/a_0)^3-\rho_{_{0\chi}}} \sqrt{\frac{-H_0^2 \Omega _{\text{BE}}+(a/a_0)^3 \rho _c-\rho_{_{0\chi}} \rho _c}{\rho _c \left((a/a_0)^3-\rho_{_{0\chi}}\right)}}-\\\nonumber&&\frac{2 \rho_{_{0\chi}} \rho _c
   \sqrt{(a/a_0)^3-\rho_{_{0\chi}}} \sqrt{\frac{\rho _c \left((a/a_0)^3-\rho_{_{0\chi}}\right)-H_0^2 \Omega _{\text{BE}}}{\rho _c \left((a/a_0)^3-\rho_{_{0\chi}}\right)}} \tanh ^{-1}\left(\sqrt{1-\frac{(a/a_0)^3
   \rho _c}{H_0^2 \Omega _{\text{BE}}+\rho_{_{0\chi}} \rho _c}}\right)}{3 \left(H_0^2 \Omega _{\text{BE}}+\rho_{_{0\chi}}\rho _c\right) \sqrt{1-\frac{(a/a_0)^3 \rho _c}{H_0^2 \Omega
   _{\text{BE}}+\rho_{_{0\chi}} \rho _c}}},
\end{eqnarray}
where $C$ is an integration constant, determined from the constraint $a=a_{_{{\rm BEC}}}$ at $t=t_{{\rm BEC}}$ as

\begin{figure*}[ht]
\centering
\includegraphics[width=3in]{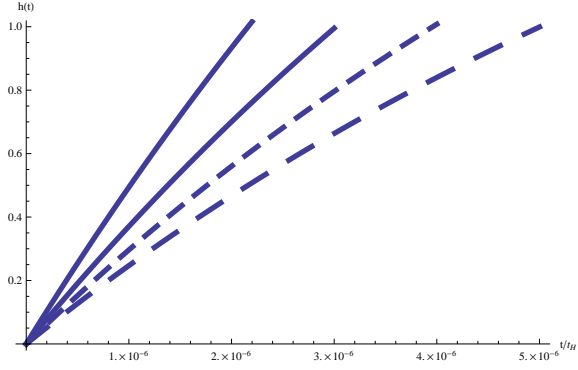}
\caption{Time evolution of the condensed dark matter fraction $h(t)$ for different values of $r$: $r=-0.15$ (solid curve), $r=-0.20$ (dotted curve), $r=-0.25$ (dashed curve) and $r=-0.30$ (long dashed curve), respectively.}
\label{stable}
\end{figure*}

\begin{figure*}[ht]
\centering
\includegraphics[width=3in]{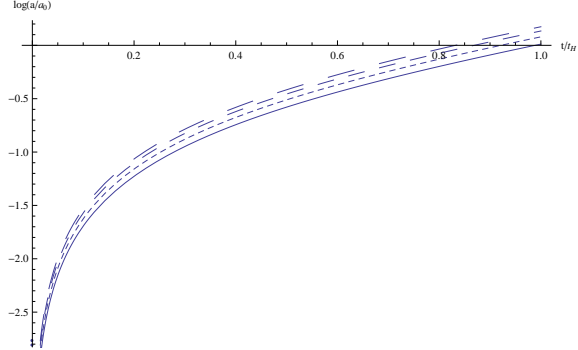}
\caption{Time evolution (in a logarithmic scale) of the scale factor of the universe in the post-Bose-Einstein condensation phase, for $a_{{\rm BEC}}=a(0)=8.1\times10^{-4}$ ($z\approx1200$), and for different values of $\rho_{0\chi}=10^{-11}$ ${\rm eV}/{\rm cm}^{3}$(solid curve), $\rho_{0\chi}=5\times10^{-11}$ ${\rm eV}/{\rm cm}^{3}$(dotted curve),  $\rho_{0\chi}=10^{-10}$ ${\rm eV}/{\rm cm}^{3}$(dashed curve), and $\rho_{0\chi}=5\times10^{-10}$ ${\rm eV}/{\rm cm}^{3}$(long dashed curve), respectively.}
\label{stable1}
\end{figure*}

\begin{eqnarray}\label{eq63}
C&=&t_{{\rm BEC}}-\frac{2}{3H_0\sqrt{\Omega_{BE}}} \sqrt{(a_{_{\rm BEC}}/a_0)^3-\rho_{_{0\chi}}} \sqrt{\frac{-H_0^2 \Omega _{\text{BE}}+(a_{_{\rm BEC}}/a_0)^3 \rho _c-\rho_{_{0\chi}} \rho _c}{\rho _c \left((a_{_{\rm BEC}}/a_0)^3-\rho_{_{0\chi}}\right)}}-\\\nonumber&&\frac{2 \rho_{_{0\chi}} \rho_c \sqrt{(a_{_{\rm BEC}}/a_0)^3-\rho_{_{0\chi}}} \sqrt{\frac{\rho _c \left((a_{_{\rm BEC}}/a_0)^3-\rho_{_{0\chi}}\right)-H_0^2 \Omega _{\text{BE}}}{\rho _c \left((a_{_{\rm BEC}}/a_0)^3-\rho_{_{0\chi}}\right)}} \tanh ^{-1}\left(\sqrt{1-\frac{(a_{_{\rm BEC}}/a_0)^3
   \rho _c}{H_0^2 \Omega _{\text{BE}}+\rho_{_{0\chi}} \rho _c}}\right)}{3 H_0\sqrt{\Omega_{BE}}\left(H_0^2 \Omega _{\text{BE}}+\rho_{_{0\chi}}\rho _c\right) \sqrt{1-\frac{(a_{_{\rm BEC}}/a_0)^3 \rho _c}{H_0^2 \Omega
   _{\text{BE}}+\rho_{_{0\chi}} \rho _c}}}
\end{eqnarray}

For the universe presented here, filled with radiation, dark energy, baryonic matter and Bose-Einstein condensed dark matter, the equation describing time evolution of the scale factor  is given by
\begin{eqnarray}\label{eq64}
\frac{1}{(a/a_{0})}\frac{d(a/a_{0})}{dt}&=&
H_{0}\sqrt{\frac{\Omega_{b,0}}{\left(\frac{a}{a_{0}}\right)^{3}}+\frac{\Omega_{{\rm rad,0}}}{\left(\frac{a}{a_{0}}\right)^{4}}+
\frac{\Omega_{{\rm BE}}}{\left(\frac{a}{a_{0}}\right)^{3}-\rho_{0,\chi}}+\Omega_{\Lambda}}\\\nonumber&&\times
\sqrt{1-\frac{H^{2}_{0}}{\sqrt{\rho_{c}}}\left(\frac{\Omega_{b,0}}{\left(a/a_{0}\right)^{3}}+
\frac{\Omega_{{\rm rad,0}}}{\left(a/a_{0}\right)^{4}}+\frac{\Omega_{{\rm BE}}}{\left(a/a_{0}\right)^{3}-\rho_{0\chi}}+\Omega_{\Lambda}\right)},~~~~~t\geq t_{{\rm BEC,}}
\end{eqnarray}
which can be integrated by considering the initial conditions $a(t_{{\rm BEC}})=a(0)=a_{{\rm BEC}}$. To study the time evolution of the scale factor in such a universe which contains BEC dark matter, it will be useful to depict it for different values of the $\rho_{0\chi}$ in Fig.$2$. It is remarkable that the existence of the condensed dark matter changes the cosmological dynamics of the universe in the post condensation phase, and the magnitude of this change decreases with the increase of the Bose-Einstein Condensate parameter $\rho_{0\chi}$.
This behavior is expected because as $\rho_{0\chi}$ increases, the corresponding
gravity contributes stronger to the dynamics and decreases the expansion
rate.
On the other hand, it is seen from Fig.2 that as $\rho_{0\chi}$ increases,
the Bose-Einstein Condensation phase (where $a=a_{{\rm BEC}}$) occurs at
earlier times. This is also justified because as $\rho_{0\chi}$ increases,
the $\rho^{{\rm BEC}}_{\chi}$ increases as well and so the scale factor  $a_{{\rm BEC}}$ corresponding
to this increased density tends toward the scale factors having higher density,
which naturally occurs at earlier times. 
\section{Conclusions}

When the Bose-Einstein condensation happens, dark matter contains two phases, the normal and the condensed phase, respectively. Under a thermodynamic condition, we expect to have the continuity of the pressure profile of these two phases right at the beginning of the condensation process. This leads to uniquely determination of dark matter density  at the condensation time as well as other thermodynamical quantities, like temperature and pressure. The explicit numerical values of the condensation quantities is related to mean square velocity of the normal dark matter $\sigma^{2}$ and   the mass $m_{\chi}$ of the dark matter particle and the scattering length $l_{a}$. As a result of the uncertainty of the values of $\sigma^{2}$, $m_{\chi}$ and $l_{a}$, apparently it is difficult to predict the precise and exact cosmological moment of the Bose-Einstein condensation and consequently the corresponding cosmological quantities. Nevertheless, by considering some ``standard'' numerical values, it will be possible to obtain a total qualitative picture of the transition. Consequently, we estimate the values of the dark matter mass and the mean velocity of the non-relativistic dark matter particles, of the order of $1\rm eV$\cite{8}, \cite{27} and $900 \rm km/s$, respectively. According to the general analysis of the condensation process, the first phase of condensation belongs to the standard $\Lambda$ cold dark matter ($\Lambda$CDM) model. In this paper, we have generalized this analysis and have gone through the loop quantum cosmology context. With the standard numerical values of the dark matter quantities, the condensation was begun at a redshift around $z=1200$. Here, we have studied the time evolution of the fraction factor $h(t)$ of the condensed dark matter through the loop quantum cosmology. In the next step, we have extracted the modified phase transition time interval in LQC, which is approximately of the order of Hubble time and shows the viability of our results. We have studied the evolution during the Bose-Einstein condensation phase
and found that this condensation process modifies the expansion rate of the universe. Also, we have studied the post-condensation phase in LQC model, describing the time evolution of universe right after the end of phase transition, which leads to (\ref{eq64}). The cosmic time dependence of the global formation of BEC dark matter (at $t_{{\rm BEC}}$) is an advantage, compared to the ordinary dark matter, which makes BEC dark matter to be preferred
than the ordinary dark matter. This is an important feature which makes BEC dark matter to be similar to the global
dark matter candidates like neutrinos, WIMPs, axions and LSP. The ordinary
dark matter has not such capability and cannot form global sources of dark
matter at a definite
cosmic time, and that is why we have favoured BEC dark matter in the present study. We have plotted the time behavior of equations (\ref{eq53}) and (\ref{eq64}) in Figures 1 and 2 for different values of $r$ and $\rho_{0\chi}$, showing the definite effect of the phase transition and Bose-Einstein condensed dark matter density in the modified cosmological history in the loop quantum cosmology.

 A major point regarding the present study is that  assuming two phases of dark matter, one as the bosonic particles before
BEC, and the other one as BEC dark matter,  ``is there any experimental indication for the
occurrence of BEC, the existence of these two phases, and the preference of one
 to the other one?'' or ``why it is
necessary to distinguish ordinary dark matter from BEC dark matter?''. Of course, at present there
is no  direct evidence for the
occurrence of BEC, the existence of these two phases, and the preference of one
 to the other one. However, in order to answer
properly these questions, we may try to answer the  alternative question that
``is there any indication for distinguishing between
the dynamics of the universe having bosonic particles without the occurrence of BEC, and the one with
BEC dark matter''. In this regard, we may point out that the time evolution of the scale factor with BEC is different from that of  without BEC for many possible reasons such as i) the equation of state parameter of BEC dark matter is different from that of bosonic particles without BEC, due to the first order phase
transition, ii) in the absence of BEC, the bosonic particles may interact
or participate in some possible transformations into other matter components, like visible matter or dark energy, each resulting in different dynamical history of the universe; however with BEC the condensed bosonic particles cannot decay into other forms of matter and they remain as the BEC dark matter permanently, with a definite dynamical
history distinguishable from variety of dynamics corresponding to the regular bosonic particles
having different  kinds of  interaction with other matter components.
\section*{Acknowledgments}
This work has been supported financially by Research Institute
for Astronomy and Astrophysics of Maragha (RIAAM) under research project
NO.1/4165-96.

\end{document}